\documentclass[a4paper]{spie}

\usepackage[]{graphicx}
\usepackage{color}

\newcommand{\ud}{\mathrm{d}}

\title{Quantum points/patterns, Part 2.\\
From quantum points to quantum patterns via multiresolution}

\author{Antonina N.~ Fedorova\supit{a} and  Michael G.~Zeitlin\supit{a}
\skiplinehalf
\supit{a}IPME RAS, V.O. Bolshoj pr., 61, 199178, St.~Petersburg, Russia
}
\authorinfo{anton@math.ipme.ru, zeitlin@math.ipme.ru,
http://www.ipme.ru/zeitlin.html,
http://mp.ipme.ru/zeitlin.html}

\begin{document} 

\begin{center}
\begin{tabular}{p{160mm}}

\begin{center}
{\bf\Large
Quantum points/patterns, Part 2.} \\
\vspace{5mm}

{\bf\Large  From quantum points to quantum}\\
\vspace{5mm}

{\bf\Large patterns via multiresolution}\\

\vspace{1cm}

{\bf\Large Antonina N. Fedorova, Michael G. Zeitlin}\\

\vspace{1cm}

{\bf\large\it
IPME RAS, St.~Petersburg,
V.O. Bolshoj pr., 61, 199178, Russia}\\
\vspace{0.2cm}
{\bf\large\it e-mail: zeitlin@math.ipme.ru}\\
\vspace{0.2cm}
{\bf\large\it e-mail: anton@math.ipme.ru}\\
\vspace{0.2cm}
{\bf\large\it http://www.ipme.ru/zeitlin.html}\\
\vspace{0.2cm}
{\bf\large\it http://mp.ipme.ru/zeitlin.html}

\end{center}

\vspace{1cm}

\begin{abstract}
It is obvious that we still have not any unified framework
covering a zoo of interpretations of Quantum Mechanics, as well as
satisfactory understanding of main ingredients of a
phenomena like entanglement. The starting point is an idea
to describe properly the key ingredient of the area, namely point/particle-like 
objects (physical quantum points/particles or, at least, structureless but quantum objects) 
and to change point (wave) functions by sheaves  to the sheaf wave functions (Quantum Sheaves).
In such an approach Quantum States are sections of the coherent sheaves 
or contravariant functors from the kinematical category describing space-time 
to other one, Quantum Dynamical Category, properly describing the complex dynamics of Quantum Patterns. 
The objects of this category are some filtrations on the functional realization 
of Hilbert space of Quantum States.
In this Part 2, the sequel of Part 1, we present a family of methods which can describe important details 
of complex behaviour in quantum ensembles:
the creation 
of nontrivial patterns, localized, chaotic, entangled or decoherent,
from the fundamental basic localized (nonlinear) eigenmodes (in contrast with orthodox gaussian-like)
in various collective models arising from the 
quantum hierarchies described by Wigner-like equations. 
\end{abstract}

\vspace{15mm}

\begin{center}
{\large Submitted to Proc. of SPIE Meeting,}\\
\vspace{0.2cm}
{\large The Nature of Light: What are Photons? IV}\\
\vspace{0.2cm}
{\large San Diego, CA, August, 2011}

\vspace{5mm}

\end{center}
\end{tabular}
\end{center}
\newpage


\maketitle

\begin{abstract}
It is obvious that we still have not any unified framework
covering a zoo of interpretations of Quantum Mechanics, as well as
satisfactory understanding of main ingredients of a
phenomena like entanglement. The starting point is an idea
to describe properly the key ingredient of the area, namely point/particle-like 
objects (physical quantum points/particles or, at least, structureless but quantum objects) 
and to change point (wave) functions by sheaves  to the sheaf wave functions (Quantum Sheaves).
In such an approach Quantum States are sections of the coherent sheaves 
or contravariant functors from the kinematical category describing space-time 
to other one, Quantum Dynamical Category, properly describing the complex dynamics of Quantum Patterns. 
The objects of this category are some filtrations on the functional realization 
of Hilbert space of Quantum States.
In this Part 2, the sequel of Part 1, we present a family of methods which can describe important details 
of complex behaviour in quantum ensembles:
the creation 
of nontrivial patterns, localized, chaotic, entangled or decoherent,
from the fundamental basic localized (nonlinear) eigenmodes (in contrast with orthodox gaussian-like)
in various collective models arising from the 
quantum hierarchies described by Wigner-like equations. 
\end{abstract} 


\section{New localized modes and patterns: why need we them?}

It is widely known that the currently available experimental techniques in the area of 
quantum physics as well 
as the present level of the understanding of phenomenological models, outstrips
the actual level of mathematical description [1], [2].
Considering the problem of describing the really existing and/or realizable  states,
one should not expect that well-known trivial states like (gaussian) coherent states, harmonic plane waves 
or eigenstates of orthodox (quantum) Hamiltonians would be enough to characterize the 
power of complex quantum phenomena. 
The complexity of a set of relevant states, including entangled (chaotic) ones 
is still far from being clearly understood and moreover from being realizable [3], [4].

Our motivations arise from the following 
general questions:

how can we represent a well localized and reasonable state in mathematically correct form?

is it possible to create entangled and other relevant states by means of these new localized 
building blocks?

The general idea is rather simple: it is well known that the
generating symmetry is the key ingredient of any modern
reasonable physical theory. 
Roughly speaking, the representation theory of the underlying 
(internal/hidden) symmetry (classical or quantum, finite 
or infinite dimensional, continuous 
or discrete) is the useful instrument for the description of (orbital) dynamics.

The proper representation theory is well known as ``local nonlinear harmonic analysis'',
in particular case of the simple underlying symmetry, affine group, aka wavelet analysis [5]--[7].
From our point of view the advantages of such approach are as follows:
\begin{itemize}
\item[\bf i)] the natural realization of localized states in any proper functional realization of 
(Hilbert) space of states,
\item[\bf ii)] the hidden symmetry of a chosen realization of the functional model describes 
the (whole) spectrum of possible states via the so-called 
multiresolution technique providing the exact multiscale decomposition.
\end{itemize}

Effects we are interested in are as follows: 
\begin{itemize}
\item[{\bf 1).}] a hierarchy of internal/hidden 
scales (time, space, phase space);
\item[{\bf 2).}] non-perturbative multiscales: 
from slow to fast contributions,
from the coarser to the finer level 
of resolu\-ti\-on/de\-composition;
\item[{\bf 3).}] the coexistence of the levels of hierarchy of multiscale dynamics with 
transitions/intermittency between scales;
\item[{\bf 4).}] the realization of the key features of the complex quantum 
world such as the existence of chaotic and/or entangled 
states with possible destruction in ``open/dissipative'' regimes due to interactions with
quantum/classical environment and transition to decoherent states.
\end{itemize}
At this level, we may interpret the effect of mysterious entanglement or ``quantum interaction''  
as a result of the simple interscale interaction or intermittency (with allusion to hydrodynamics),
i.e. the mixing of orbits generated by multiresolution representation of the hidden underlying symmetry.
Surely, the existence of such a symmetry is a natural physical property 
of the model as well as the structure/type of the space of representation and its proper functional 
realization.
So, instantaneous quantum interaction 
materializes not in the physical space-time variety 
but in the space of the representation of hidden symmetry
along the orbits/scales constructed by proper representations. 
Such an approach provides the explicit analytical construction for solutions of
c- and q-hierarchies and their important reductions starting from the quantization 
of c-BBGKY hierarchy [8]--[22].
It is based 
on tensor algebra extensions of multiresolution
representation [5] for states and observables and variational formulation [8]--[22].
We provide the explicit representation for the hierarchy of n-particle 
reduced distribution functions 
in the base of the
high-localized generalized coherent (regarding underlying generic symmetry 
(affine group in the simplest case)) 
states given by the polynomial tensor algebra of proper (in exact sense) basis functions 
(wavelet families, wavelet packets [6]--[7]), which 
takes into account
contributions from all underlying hidden multiscales
from the coarsest scale of resolution to the finest one to
provide the full information about (quantum) dynamical process.
The difference between classical and quantum case is concentrated
in the structure of the set of operators and proper functional spaces where they are realized,
included in the set-up, and, of course,
depends on the method of quantization.
But, in the naive Wigner-Weyl approach for the quantum case, the symbols of operators  
play the same role as usual functions in the classical case. 
In some sense,
our approach for ensembles (hierarchies) resembles Bogolyubov's one and related approaches 
but we do not use any perturbation technique (like virial expansion)
or linearization procedures.
Most important, that 
numerical modeling in all cases shows the creation of
various internal (coherent)
structures from localized modes, which are related to the (meta)stable (equilibrium) or 
unstable type of behaviour and corresponding patterns (waveletons) formation [8]--[22].

We start from the second quantized 
representation for an algebra of observables 
$A=(A_0,A_1,\dots,A_s,...)$
in the standard form
\begin{eqnarray*}
&&A=A_0+\int dx_1\Psi^+(x_1)A_1\Psi(x_1)+\dots\\
&&+(s!)^{-1}\int dx_1\dots dx_s\Psi^+(x_1)\dots
\Psi^+(x_s)A_s\Psi(x_s)\dots \Psi(x_1)+\dots.
\end{eqnarray*}

N-particle Wig\-ner functions
allow to consider them as partitions representing some useful quasiprobabilities. 
The full description for quantum ensemble can be done by the hierarchy
of functions (symbols):

$$
W=\{W_s(x_1,\dots,x_s), s=0,1,2\dots\},\\
$$
which are solutions of Wigner (pseudodifferential) equations:

\begin{eqnarray}
\frac{\partial W_n}{\partial t}=-\frac{p}{m}\frac{\partial W_n}{\partial q}+
\sum^{\infty}_{\ell=0}\frac{(-1)^\ell(\hbar/2)^{2\ell}}{(2\ell+1)!}
\frac{\partial^{2\ell+1}U_n(q)}{\partial q^{2\ell+1}}
\frac{\partial^{2\ell+1}W_n}{\partial p^{2\ell+1}}.
\end{eqnarray}
The similar Lindblad equations describe the important decoherence processes [4].

\section{Variational multiresolution representation}

We obtain our multiscale/multiresolution representations for solutions of Wig\-ner-like equations
(1) via the variational--multiresolution approach. 
We represent the solutions as 
decomposition into localized eigenmodes  
related to a underlying set of scales corresponding to proper orbits generated 
by action of hidden internal symmetry, like (non-abelian) 
affine group in the simplest but important case of wavelet analysis: 
$$
W_n(t,q,p)=\displaystyle\bigoplus^\infty_{i=i_c}W^i_n(t,q,p),
$$
where value $i_c$ corresponds to the coarsest level of resolution
$c$ in 
the full Multiresolution Analysis Decomposition (MRA)
of the underlying functional space [5]:
$$
V_c\subset V_{c+1}\subset V_{c+2}\subset\dots
$$
and $p=(p_1,p_2,\dots),\quad q=(q_1,q_2,\dots),\quad x_i=(p_1,q_1,\dots,p_i,q_i)$ 
are coordinates in phase space.
We introduce the Fock-like space structure on the whole space of internal hidden scales
$$
H=\bigoplus_i\bigotimes_n H^n_i
$$
for the set of n-partial Wigner functions (states):
$$
W^i=\{W^i_0,W^i_1(x_1;t),\dots,
W^i_N(x_1,\dots,x_N;t),\dots\},
$$
where
$W_p(x_1,\dots, x_p;t)\in H^p$,
$H^0=C,\quad H^p=L^2(R^{6p})$ or any different proper functional spa\-ce 
with the natural Fock space like norm: 
$$
(W,W)=W^2_0+
\sum_{i}\int W^2_i(x_1,\dots,x_i;t)\prod^i_{\ell=1}\mu_\ell.
$$
First of all, we consider $W=W(t)$ as a function of time only,
$W\in L^2(R)$, via
multiresolution decomposition which naturally and efficiently introduces 
an infinite sequence of the underlying hidden scales [5].
We have the contribution to
the final result from each scale of resolution from the whole
infinite scale of spaces or more correctly mathematically, from the Tower of Filtration.
The closed subspace
$V_j (j\in {\bf Z})$ corresponds to  the level $j$ of resolution
and satisfies
the following properties:
let $D_j$ be the orthonormal complement of $V_j$ with respect to $V_{j+1}$: 
$
V_{j+1}=V_j\bigoplus D_j.
$
Then we have the following decomposition:

$$
\{W(t)\}=\bigoplus_{-\infty<j<\infty} D_j 
=\overline{V_c\displaystyle\bigoplus^\infty_{j=0} D_j},
$$

in case when $V_c$ is the coarsest scale of resolution.
The subgroup of translations generates a basis for the fixed scale number:
$
{\rm span}_{k\in Z}\{2^{j/2}\Psi(2^jt-k)\}=D_j.
$
The whole basis is generated by the action of the full affine group:
$$
{\rm span}_{k\in Z, j\in Z}\{2^{j/2}\Psi(2^jt-k)\}=
{\rm span}_{k,j\in Z}\{\Psi_{j,k}\}
=\{W(t)\}.
$$
After the construction of the multidimensional tensor product bases [4],  
the next key point is 
the so-called Fast Wavelet Transform (FWT) [5]--[7], 
demonstrating that for a large class of
operators the wavelet functions are a good 
approximation for true eigenvectors and the corresponding 
matrices are almost diagonal. 
We have the simple linear para\-met\-rization of the
matrix representation of  our operators in the localized wavelet bases
and of the action of
these operators on arbitrary vectors/states in the proper functional space.
FWT provides  the maximum sparse and useful form for the wide classes 
of operators [5]--[7].
After that,
we can obtain our multiscale/mul\-ti\-re\-so\-lu\-ti\-on 
representations for observables (symbols), states, partitions
via the variational approaches.

Let $L$ be an arbitrary (non)li\-ne\-ar dif\-fe\-ren\-ti\-al\-/\-in\-teg\-ral operator 
 with matrix dimension $d$
(finite or infinite), 
which acts on some set of functions
from $L^2(\Omega^{\otimes^n})$:  
$\quad\Psi\equiv\Psi(t,x_1,x_2,\dots)=\Big(\Psi^1(t,x_1,x_2,\dots), \dots$,
$\Psi^d(t,x_1,x_2,\dots)\Big)$,
 $\quad x_i\in\Omega\subset{\bf R}^6$, $n$ is a number of particles:
\begin{eqnarray*}
L\Psi&\equiv& L(Q,t,x_i)\Psi(t,x_i)=0,\\
Q&\equiv& Q_{d_0,d_1,d_2,\dots}(t,x_1,x_2,\dots,
\partial /\partial t,\partial /\partial x_1,
\partial /\partial x_2,\dots,
\int \mu_k)
\\
&=&\sum_{i_0,i_1,i_2,\dots=1}^{d_0,d_1,d_2,\dots}
q_{i_0i_1i_2\dots}(t,x_1,x_2,\dots)
\Big(\frac{\partial}{\partial t}\Big)^{i_0}\Big(\frac{\partial}{\partial x_1}\Big)^{i_1}
\Big(\frac{\partial}{\partial x_2}\Big)^{i_2}\dots\int\mu_k.
\end{eqnarray*}
Let us consider the $N$ mode approximation:
$$
\Psi^N(t,x_1,x_2,\dots)=
\sum^N_{i_0,i_1,i_2,\dots=1}a_{i_0i_1i_2\dots}
 A_{i_0}\otimes 
B_{i_1}\otimes C_{i_2}\dots(t,x_1,x_2,\dots)
.
$$
We will determine the expansion coefficients from the following conditions
(related to the proper choosing of variational approach) which are nothing 
but Generalized Dispersion Relations (GDR) :
\begin{eqnarray}
&&\ell^N_{k_0,k_1,k_2,\dots}\equiv 
\int(L\Psi^N)A_{k_0}(t)B_{k_1}(x_1)C_{k_2}(x_2)\ud t\ud x_1\ud x_2\dots=0.
\end{eqnarray}
Thus, we have exactly $dN^n$ algebraical equations for  $dN^n$ unknowns 
$a_{i_0,i_1,\dots}$.
This variational ap\-proach reduces the initial problem 
to the problem of solution 
of functional equations at the first stage and 
some algebraical problems at the second one.
It allows to unify the multiresolution expansion with variational 
construction [8]--[22]. 
As a result, the solution is parametrized by the solutions of two sets of 
reduced algebraical
problems, one is linear or nonlinear
(depending on the structure of the generic operator $L$) and the rest are linear
problems related to the computation of the coefficients of reduced 
algebraic equations. It is also related to the choice of exact measure of localization
(including the class of smoothness), which is proper for our set-up.
These coefficients can be found  via functional/algebraic methods
by using the
compactly supported wavelet basis or any other wavelet families [6]--[7].
As a result, the solution of the hierarchies as in c-
as in q-region, has the 
following mul\-ti\-sca\-le or mul\-ti\-re\-so\-lu\-ti\-on decomposition via 
nonlinear lo\-ca\-li\-zed eigenmodes (Fig.~1): 
{\setlength\arraycolsep{0pt}
\begin{eqnarray}
W(t,x_1,x_2,\dots)&=&
\sum_{(i,j)\in Z^2}a_{ij}U^i\otimes V^j(t,x_1,\dots),\nonumber\\
V^j(t)&=&
V_N^{j,slow}(t)+\sum_{l\geq N}V^j_l(\omega_lt), \ \omega_l\sim 2^l,\\ 
U^i(x_s)&=&
U_M^{i,slow}(x_s)+\sum_{m\geq M}U^i_m(k^{s}_mx_s), \ k^{s}_m\sim 2^m,\nonumber
\end{eqnarray}
}
which corresponds to the full multiresolution expansion in all underlying time/space 
scales.
The formulas (3) give the expansion into a slow part
and fast oscillating parts for arbitrary $N, M$.  So, we may move
from the coarse scales of resolution to the 
finest ones for obtaining more detailed information about the dynamical process.
In this way, one obtains contributions to the full solution
from each scale of resolution or each time/space scale or from each nonlinear eigenmode.
It should be noted that such representations 
give the best possible localization
properties in the corresponding (phase)space/time coordinates. 
Representation (3) do not use perturbation
techniques or linearization procedures.
Numerical calculations are based on compactly supported
wavelets and wavelet packets and on the evaluation of  
accuracy on 
the level $N$ of the corresponding cut-off of the full system 
regarding Fock-like norm described above:
$$
\|W^{N+1}-W^{N}\|\leq\varepsilon.
$$

\begin{figure}[b]
\begin{center}
\includegraphics*[width=50mm]{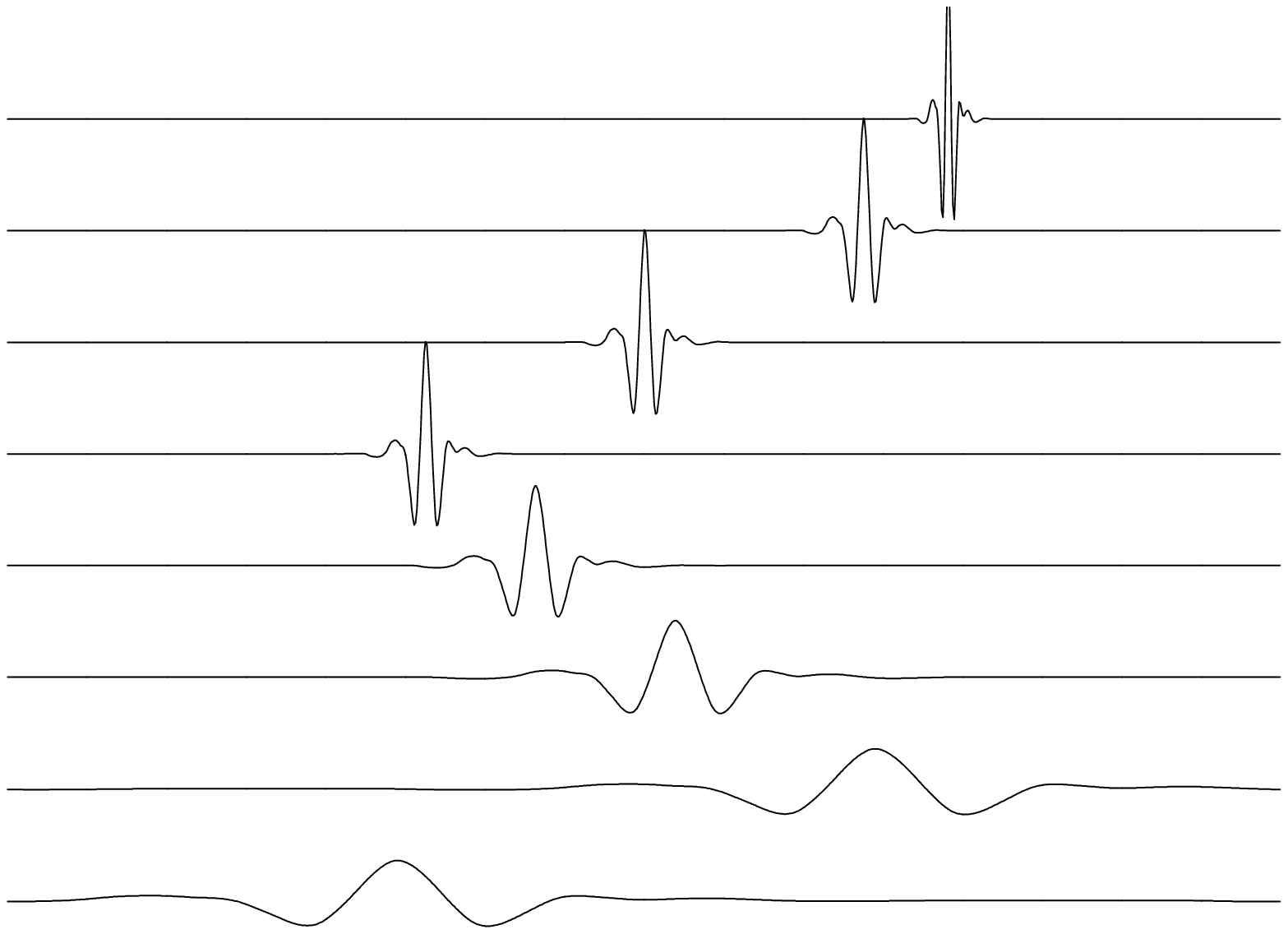}
\end{center}
\caption{Nonlinear localized basis eigenmodes.}
\end{figure}

\begin{figure}[t]
\begin{center}
\includegraphics*[width=110mm]{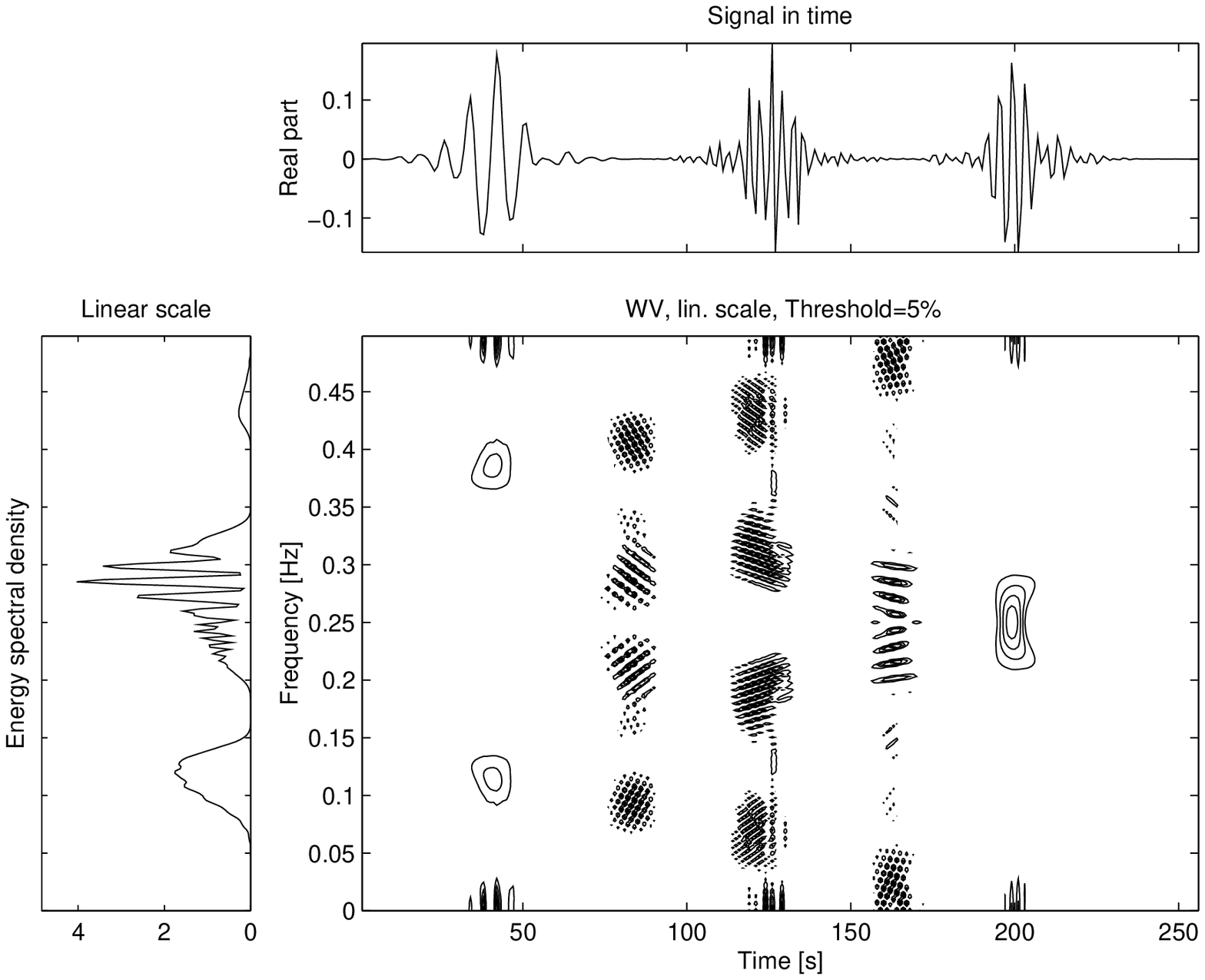}
\end{center}
\end{figure}
\begin{figure}[h]
\begin{center}
\includegraphics*[width=110mm]{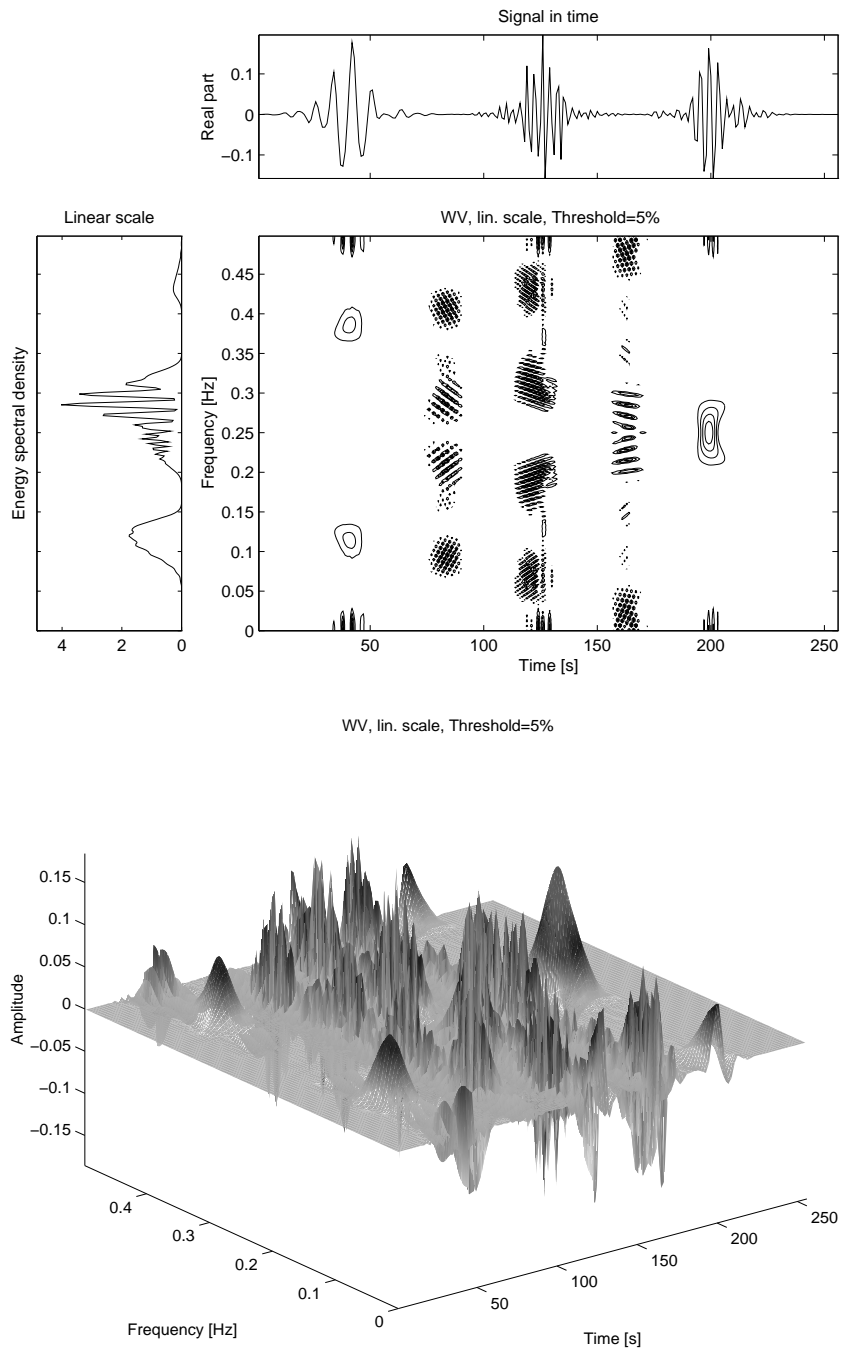}
\end{center}
\caption{Wigner function for three wavelet packets: direct modeling}
\end{figure}

\newpage

\begin{figure}
\begin{center}
\includegraphics*[width=70mm]{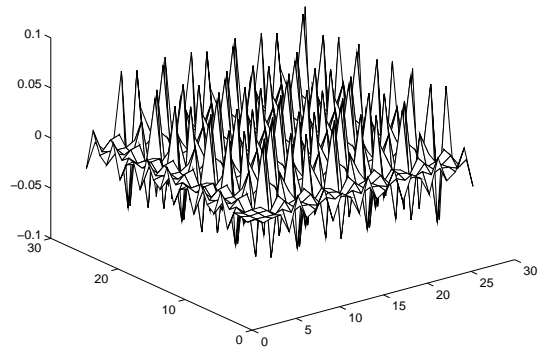}
\end{center}
\caption{MRA approximation for Wigner function.}
\end{figure}

\begin{figure}
\begin{center}
\includegraphics*[width=70mm]{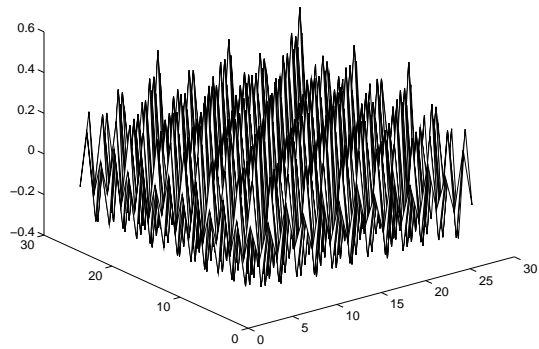}
\end{center}
\caption{MRA approximation for Wigner function: chaotic-like quantum pattern.}
\end{figure}

\begin{figure}
\begin{center}
\includegraphics*[width=70mm]{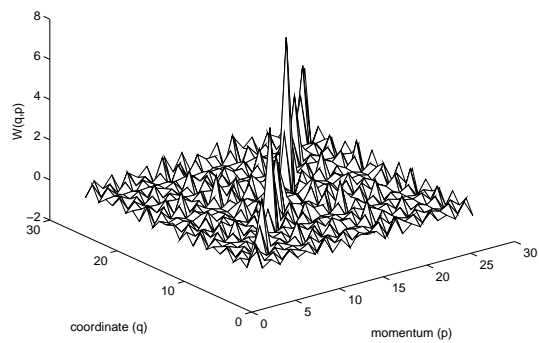}
\end{center}
\caption{Localized quantum pattern: (waveleton) Wigner function.}
\end{figure}

\begin{figure}
\begin{center}
\includegraphics*[width=70mm]{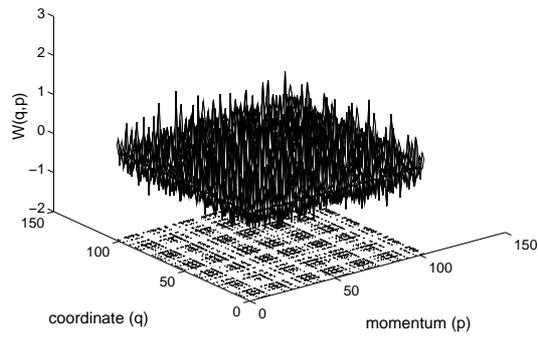}
\end{center}
\caption{Entangled-like Wigner function.}
\end{figure}

\begin{figure}
\begin{center}
\includegraphics*[width=70mm]{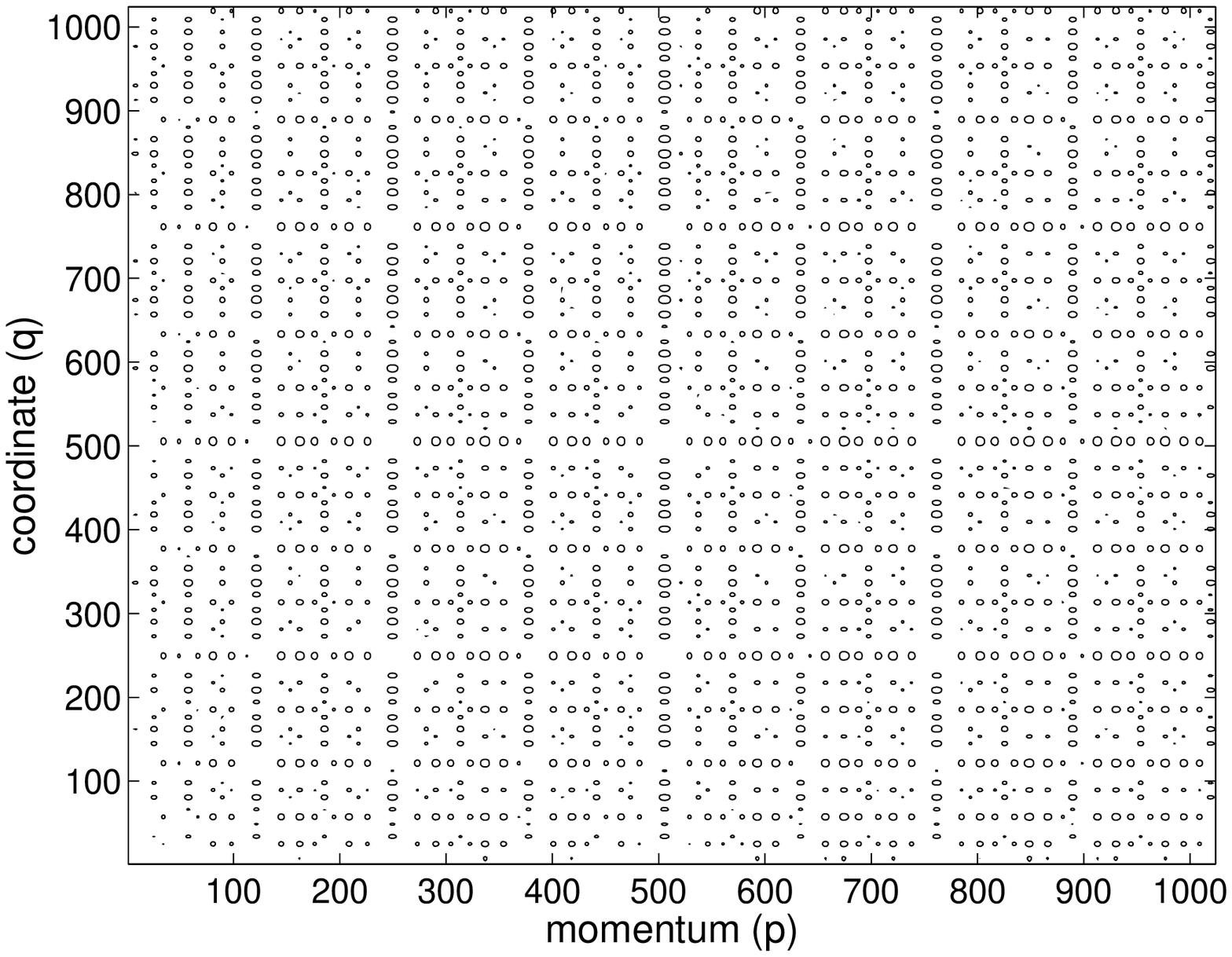}
\end{center}
\caption{Interference picture at the scale level four: approximation for Wigner function.}
\end{figure}

\begin{figure}
\begin{center}
\includegraphics*[width=70mm]{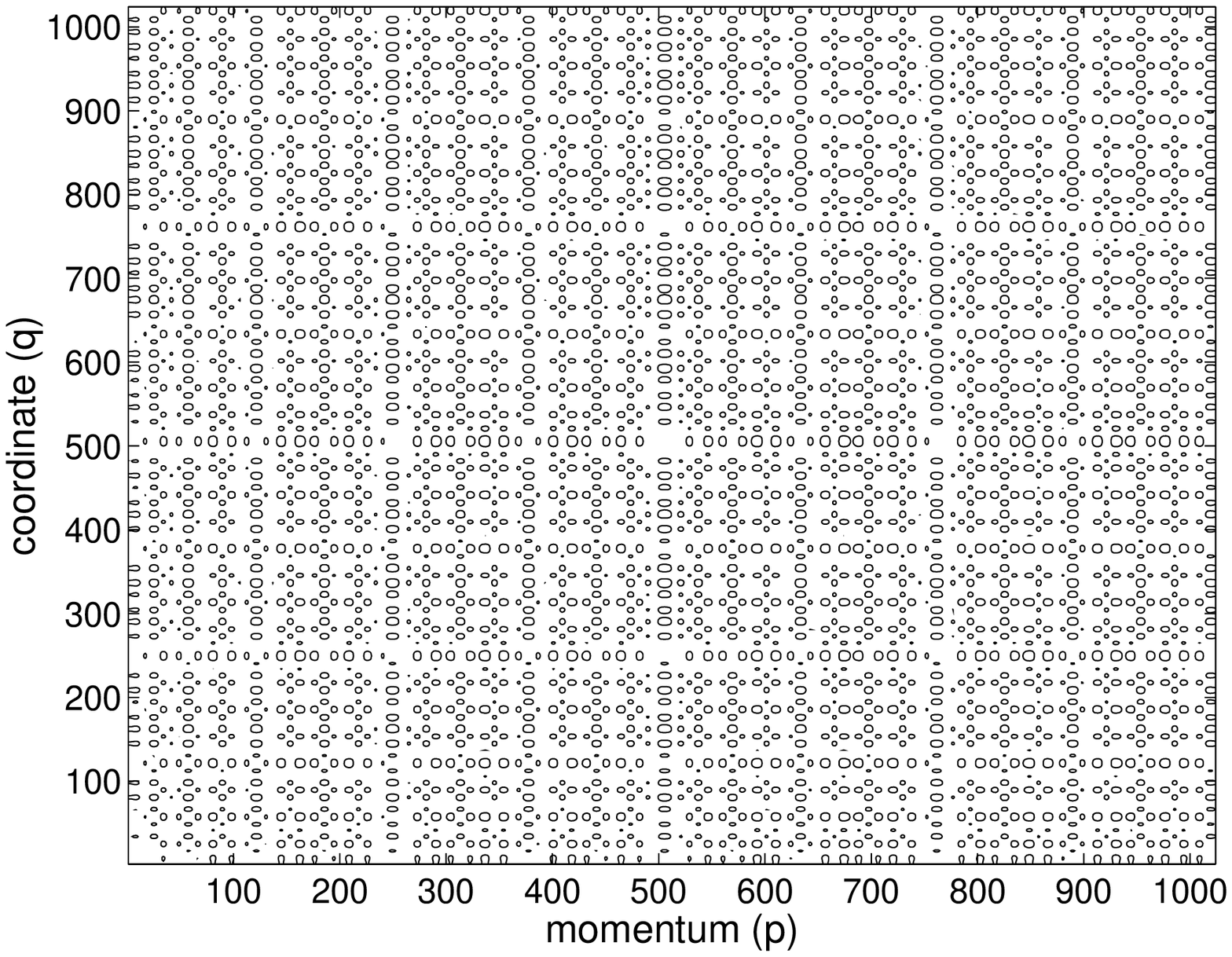}
\end{center}
\caption{Interference picture at the scale level six: approximation for Wigner function.}
\end{figure}

\section{Conclusions}
By using high localized nonlinear eigenmodes with their best phase space      
localization  properties, we can describe the full zoo of possible complex patterns generated
from localized (coherent) structures/orbits in      
quantum systems with complicated behaviour due to process of quantum self-organization (Figs.~2--8).

The numerical simulation demonstrates the formation of various (meta) stable patterns or orbits 
generated by internal hidden symmetry from generic 
high-localized fundamental modes (Fig.~1).
These (nonlinear) eigenmodes, definitely, are more realistic for the modeling of 
classical/quantum dynamical process  than infinite smooth linear gaussian-like
coherent states. 
Here we mention only the best convergence properties of the expansions 
based on wavelet packets, which  realize the minimal Shannon entropy property
and the exponential control of the convergence of expansions like (3).

Fig.~2 demonstrates results of direct 
modeling for the non-trivial Wigner function with non-trivial interference picture 
for three best localized 
wavelet packets (Fig.~1) [6], [7].

Fig.~5 presents waveleton state defined as a state with minimum entropy and zero measure, which is
generated by a finite number of fundamental modes only (more exactly, 
only a few modes contribute to the energy spectrum). 
It corresponds to the (possible) result of 
einselection [4] after decoherence process started from chaotic/entangled-like state (Figs.~4,~6).

Figs.~3, 4 and 7, 8 demonstrate the steps of multiscale resolution from level four to level six, 
or the degrees of interference, or degree of self-interaction, or intermittency-like behaviour 
during the quantum interaction/evolution   
of entangled states or quantum self-organization leading to the growth of the degree of entanglement.

It should be noted that, in addition,
we can control the type of behaviour on the level of the reduced algebraic 
system (Generalized Dispersion Relation) (2). We hope that it will be important in practical applications.
Refs. [8]--[23] contains a lot of related methods and approaches in similar complex physical problems.

\section{Summary and perspectives}

In these two Parts we considered some generalization of the theory of quantum states, which
is based on the analysis of long standing problems and
unsatisfactory situation with  possible interpretations
of quantum mechanics. We demonstrate that the consideration of quantum
states as sheaves
can provide, in principle, more deep understanding of some phenomena.
The key ingredients of the proposed construction are the families of
sections of sheaves with values
in the category of the functional realizations of infinite-dimensional
Hilbert
spaces with special (multiscale) filtration.

The questions we hope to answer are:
\begin{itemize}
\item[\bf i)] How may we enlarge the amount of (physical) information stored in one
(quantum) physical point?

\item[\bf ii)] Relation between structureless geometrical points and physical points
(or point objects like (point) particles)
    with rich (possible hidden) structure.

\item[\bf iii)] How we may ``resolve'' (physical) point/quantum state to provide such a
structure.

\item[\bf iv)] A new look in new framework for localization, entanglement, measurement,
and all that.

\item[\bf v)] How we may to explain/re-interpret a standard zoo of standard
interpretations/phenomena (multiverse,
wave functions collapse, hidden parameters, Dirac self-interference,
ensemble interpretation, etc, etc.)
of Quantum Mechanics in the new framework.
\end{itemize}

The long-range aims of approaches presented in Part 1 [23] are to compare the following key
objects which are basic for any type of the exposition of Quantum
Mechanics (and other related areas):

{\it Geometrical points vs. Physical Points (or Point Objects, or
One-Point-Patterns);

Point functions vs. Sheaves;

Partial Differential Equations (and proper orthodox approaches)

vs.

Pseudodifferential Equations (via Microlocal analysis and all that).}

The more heuristic Part 1 [23] is continued by Part 2 where we make a sketch of technical details and comparison:

{\it Fourier/Gaussian modes vs. (pretty much) Localized (but non-gaussian)
Physical Modes;

Fourier Analysis vs. Local Nonlinear Harmonic Multiscale Analysis (including
wave- and other -lets and multiresolution);

Standard Quantum Images vs. Quantum Patterns generated by Multiresolution};

Definitely, the final point after unification of the constructions from both Parts is

{\it Categorification Procedure(s) for Quantum Mechanics (QM).}

It is more or less obvious that we do not have any unified framework
covering a zoo of (mostly) discrepant interpretations of QM, as well as
satisfactory explanation/understanding of  main  ingredients of a
phenomena like entanglement, etc.

The starting point is an idea to describe (to resolve)
the key object of the area, namely (quantum) point
objects (quantum physical points or particles, e.g., photons or electrons, 
structureless at first glance or structuredness at all, in the physical reality).

Usually, for the modeling of real
physical point objects, one can consider
equivalence between them and standard geometrical points (at the moment,
the concrete description of the set to which such points belong as
one-element subsets, does not matter).

As direct consequence, our dynamical variables like wave function or
density
matrix or the Wigner function are described by means of point functions or
what
mathematicians mean by standard functions.
But, as we can understand, a geometrical point is structureless and it
seems that we need much more to enlarge the amount of data/information
corresponding to this generic object.
To advocate this Hypothesis in the present context it is worth noting
 Dirac's famous sentence
"an electron can interact only itself via the process of quantum
inteference".
Roughly speaking, from this perspective it means  that a "point particle"
needs and must have a non-trivial complicated structure.
It seems reasonable to have the rich structure for a
model of the (quantum) physical point (usually named as "particle") in
comparison
with the structureless geometrical point.
All above looks like Physical Hypothesis and it may be not so clear but at
the same
time it is more or less well known  from the mathematical point of view if
we
accept the right (Wigner-Moyal-$\star$--Quantization) picture for description of Quantum World.
In that framework, more exactly Strict Deformation Quantization approach,
all
equations are pseudodifferential and as immediate consequence
we need to change point functions by sheaves what provides the clear
resolution of
the (Physical) Point: as a result it is no more structureless but acquires
the rich (possible
hidden at first glance) structure.

To summarize, our first three Hypotheses are as follows:

\noindent {\bf (Physical) Hypothesis 1}

Physical Point Object (physical point, point particle) is not a
structureless
object and cannot be described by means of the geometrical point (in the
standard math
sense). Instead of that, Physical Points have a rich (infinite) hidden structure.

\noindent{\bf (Physical version) Hypothesis 2}

Quantum Dynamical Variables (wave functions, density matrix, Wigner
functions)
are not point functions but sheaves defined on the properly chosen
space-time manifold (or topological space, or variety, or even scheme).

\noindent{\bf (Math version) Hypothesis 2}

To provide the structuredness of the Physical Point, allowing
to enlarge a number of useful properties and increase the amount of
corresponding data inside, we consider it together with
a proper generalization of wave function as a section/fiber of proper sheaf
(in a proper category of objects) defined on a proper model (category) of
space-time.

\noindent{\bf (Physical) Hypothesis 3}

Deformation Quantization Picture (roughly speaking, Wigner-Weyl-Moyal)
is not in contradiction with Hypotheses 1 and 2 and allow us to consider the
proper
description.

\noindent{\bf (Math) Hypothesis 3}

Microlocal Analysis of proper Pseudodifferential Dynamical Equations (like
the Wigner one) allows to create a model of infinite hierarchy for
the correct representation of Physical Point.

In such an approach {\bf Quantum States}
are (roughly speaking) sections of the so-called {\bf coherent sheaves} or
{\bf contravariant functors} from the proper category corresponding to space-time to
other one properly
describing the complex dynamics of Quantum States/Patterns. As we sketched
in this Part 2, the objects of this category are some filtrations 
on the proper functional realization of Hilbert
space of States.
In this picture, a result of {\bf measurement} corresponds to the so called {\bf direct (inductive)
limit}.

Definitely, we need a proper analytical/numerical machinery to realize all that.
It is exactly the subject of approaches presented in Part 2.
It is not worth mentioning that all above is well known in Mathematics long
ago. But we hope to convert this knowledge to a physical acceptable and
computational form. We present additional details in a separate paper [23].

\acknowledgements

We are very grateful to Prof. Chandrasekhar Roychoudhuri and Michael 
Ambroselli (University of Connecticut, Storrs),
and Matthew Novak (SPIE) for their kind 
attention and help provided our presentations during 
SPIE2011 Meeting ``The Nature of Light: What are Photons? IV''
at San Diego. 
We are indebted to Dr. A. Sergeyev (IPME RAS/AOHGI) for his encouragement.

\end{document}